\documentclass[12pt]{article}

\usepackage{scicite}

\usepackage{times}

\usepackage{amsmath}
\usepackage{graphicx}
\usepackage{amssymb}
\usepackage{color}
\usepackage{ulem}
\usepackage{braket}

\newcommand{\ii}{{\rm i}}
\newcommand{\ee}{{\rm e}}

\newenvironment{sciabstract}{%
\begin{quote} \bf}
{\end{quote}}

\title{Topological Origin of Equatorial Waves}

\author{Pierre Delplace$^{1\ast}$ J. B. Marston$^{2\ast}$ Antoine Venaille$^{1\ast}$\\
\\
\normalsize{$^{1}$Univ Lyon, Ens de Lyon, Univ Claude Bernard, CNRS, Laboratoire de Physique}\\
\normalsize{AF-69342 Lyon, France}\\
\normalsize{$^{2}$ Department of Physics, Box 1843, Brown University, Providence, RI 02912-1843 USA.}\\
\normalsize{$^{\ast}$ E-mails:  pierre.delplace@ens-lyon.fr, marston@brown.edu, antoine.venaille@ens-lyon.fr}
}

\date{}

\begin{document} 


\baselineskip24pt


\maketitle


\begin{sciabstract}


Topology sheds new light on the emergence of unidirectional edge waves in a variety of physical systems, from condensed matter to artificial lattices.  Waves observed in geophysical flows are also robust to perturbations, which suggests a role for topology.  We show a topological origin for two celebrated equatorially trapped waves known as Kelvin and Yanai modes, due to the Earth's rotation that breaks time-reversal symmetry. The non-trivial structure of the bulk Poincar\'e wave modes encoded through the first Chern number of value $2$ guarantees existence for these waves. This invariant demonstrates that ocean and atmospheric waves share fundamental properties with topological insulators, and that topology plays an unexpected role in the Earth climate system.

One sentence summary: Equatorial Kelvin and Yanai waves, important components of Earth's climate system, have an origin in topology.

\end{sciabstract}


Symmetries and topology are central to an understanding of physics. In condensed matter,  topology explains the precise quantization of the Hall effect \cite{Thouless:1982kq}, where a magnetic field breaks the discrete symmetry of time reversal.  Interest in topological  properties was reinvigorated following the discovery of the quantum spin Hall effect, and the subsequent classification of different  states of matter according to discrete symmetries \cite{Hasan:2010ku}. Recently topologically protected edge excitations have been found in artificial lattices of various types \cite{goldman2016topological,lu2014topological,huber2016topological}. A correspondence between topological properties of waves in the bulk and the existence of unidirectional edge modes along boundaries exists in all these systems~\cite{hatsugai1993chern,fukui}. {The edge modes fill frequency or energy gaps found in the bulk and are immune to various types of disorder. We show that topologically protected edge waves also manifest in atmospheres and oceans.} 

Equatorial Kelvin and mixed Rossby-gravity (Yanai) waves are edge modes that propagate energy along the Earth's equator with eastward group velocity \cite{vallis2017atmospheric}.  Remarkably, the dispersion relations for these waves (Figure 1a) were derived within the framework of the rotating shallow water model~\cite{Matsuno:1966wt} just prior to their first observation in the 1960s. Since then, observations of the atmosphere have revealed a robust signature of these trapped modes in wavenumber-frequency spectra~\cite{Kiladis2009} (Figure 1b). Equatorial Kelvin and Yanai waves have been shown to play a crucial role in several aspects of climate dynamics. For instance, {Kelvin waves are a key component of the El Ni\~no Southern Oscillation, traveling across the waters of the Pacific ocean~\cite{miller1988geosat}.  The waves are also part of the quasi-biennal oscillation in the stratosphere, and are thought to be an important component of the Madden Julian Oscillation in the troposphere~\cite{zhang2005madden}.

\begin{figure}[ht]
\center
\includegraphics[height=7.cm]{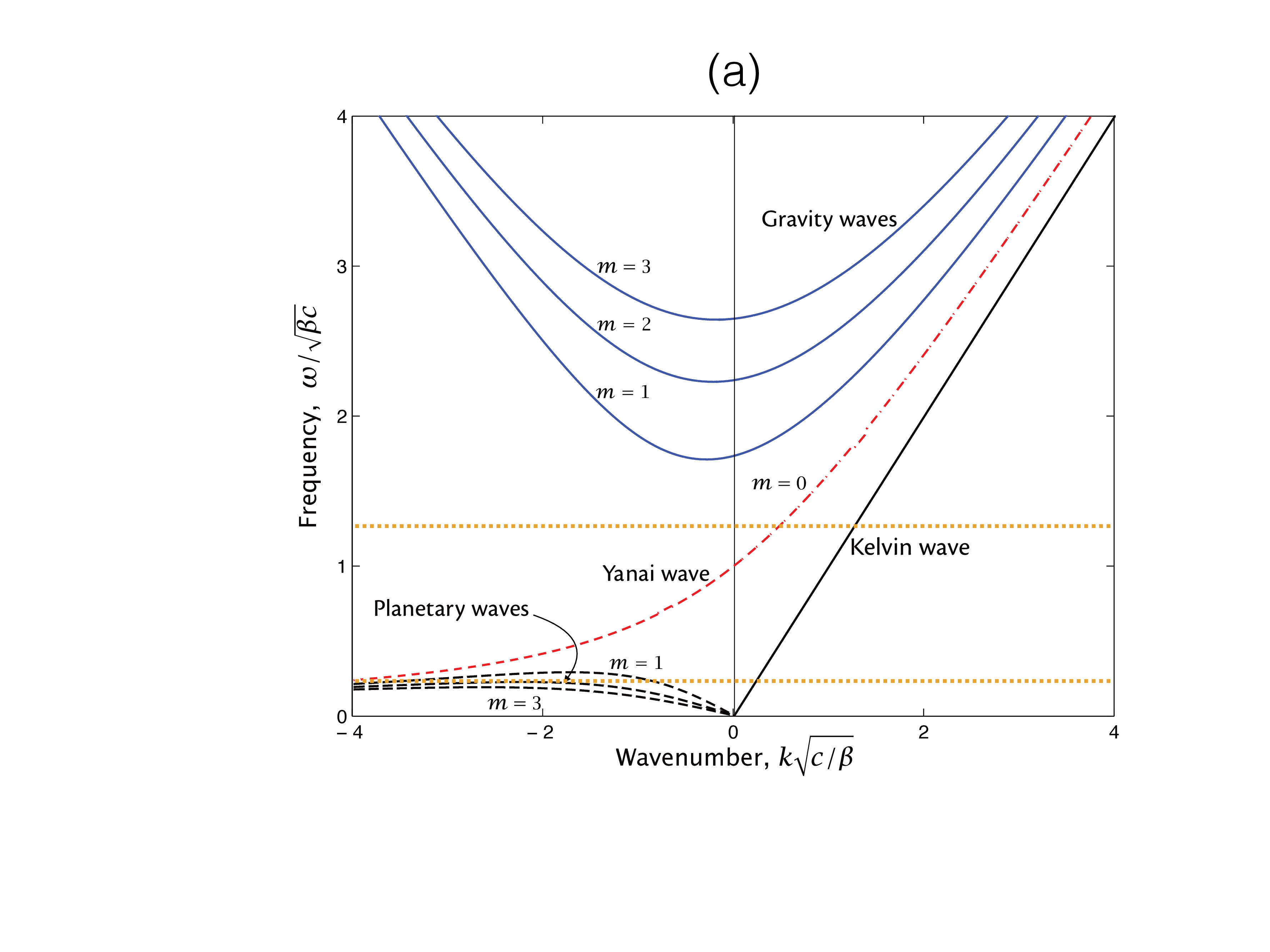}\includegraphics[height=7cm]{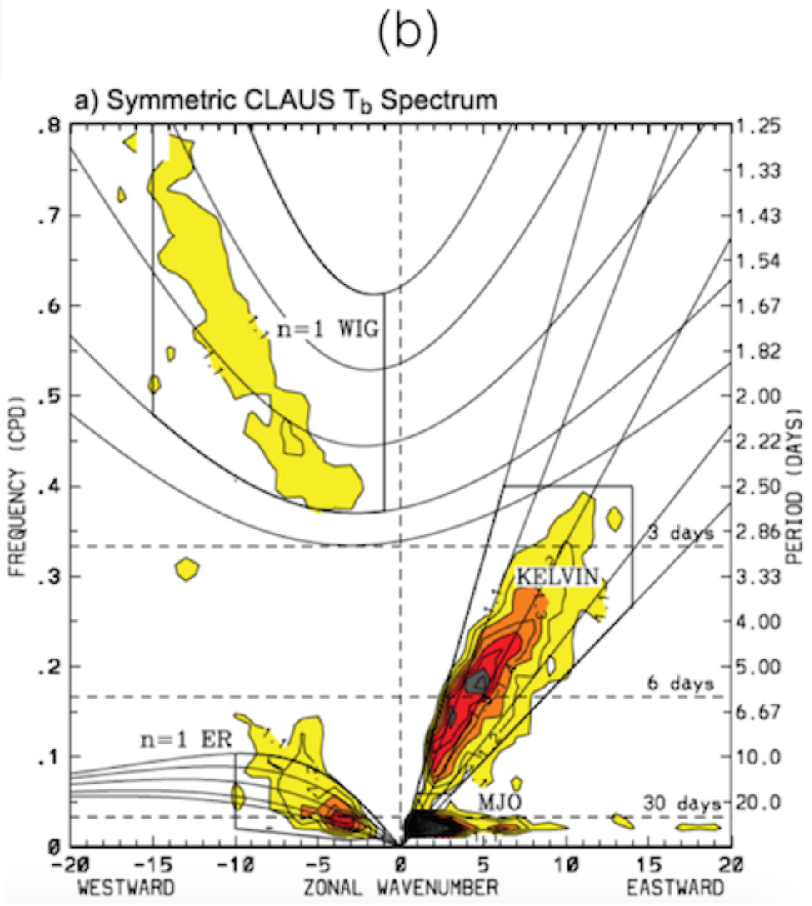} 

\caption{Dispersion spectrum of equatorial waves.  (a) Dispersion relation for shallow water waves on an equatorial beta plane with linear variations of the Coriolis parameter with the latitude ($f=\beta y$).  The dispersion relation for negative frequencies is obtained by symmetry with respect to the origin $(k_x=0,\omega=0)$. The frequency gap between low frequency planetary (Rossby) waves and high frequency inertia-gravity (Poincar\'e) waves is filled by two modes with eastward group velocity, namely the equatorial Kelvin and mixed Rossby-gravity (Yanai) waves.  Horizontal dotted orange lines indicate the intermediate, and low, frequency wave trains used in the scattering simulations of \cite{supp}.  Adapted from \cite{vallis2017atmospheric}.   
(b) Observational evidence for the appearance of the Kelvin mode in frequency-wavenumber spectra of the atmosphere.  The component that is symmetric with respect to reflection about the equator is shown.  Reproduced from~\cite{Kiladis2009}.}
\label{figure1}
\end{figure}

The fact that Yanai and Kelvin waves are equatorially trapped unidirectional modes filling a frequency gap between  low frequency planetary (Rossby) wave band and the high frequency  inertia-gravity (Poincar\'e) wave band \cite{vallis2017atmospheric}, as shown in Figure 1a, suggests they can be interpreted as topological boundary states, similar to those emerging in various topological insulating media. More precisely, bulk (Poincar\'e and/or Rossby) waves possess a  topological property, that should be directly related to the existence of these two unidirectional boundary waves, by virtue of the bulk-boundary correspondence~\cite{hatsugai1993chern,fukui}. According to this correspondence, the number of states inherited by a band when the zonal (directed along the equator) wavenumber $k_x$ varies from $-\infty$ to $+\infty$ is given by an integer-valued topological number called the first Chern number. The first Chern number is an integer that quantifies the number of phase singularities in a bundle of eigenmodes parameterized on a closed manifold. These singularities are somewhat analogous to amphidromic points ($\pm2\pi$ phase vortices of tidal modes), but they occur in parameter space rather than in physical space. We demonstrate the existence of a non-trivial global structure in the bulk Poincar\'e modes as being encoded through the first Chern number of value $\pm2$, thus ensuring the existence of $2$ unidirectional edge modes at the equator that fill the two frequency gaps, in agreement with the existence of Kelvin and Yanai waves. The existence of the frequency gap originates from a broken time-reversal symmetry of the flow model due to Earth's rotation. The structure of tidal modes~ \cite{berry1987bakerian} and bifurcations in large scale geophysical flow~\cite{gallet2012reversals} have previously invoked the effect of breaking time-reversal symmetry.  Our study shows that another far reaching consequence of this broken symmetry is to confer non-trivial topological properties to bundles of fluid waves, giving rise to robust edge states.

The rotating shallow water equations \cite{vallis2017atmospheric} that describe the dynamics of a thin layer of fluid on a two-dimensional surface of height $h(\mathbf{x},t)$ and horizontal velocity $\mathbf{u}(\mathbf{x},t)$ furnish a minimal model for equatorial waves:
\begin{eqnarray}
\partial_t h +\nabla \cdot \left (h\mathbf{u}\right) &=& 0,\\
\partial_t \mathbf{u} + (\mathbf{u} \cdot \mathbf{\nabla}) \mathbf{u} &=& -g \mathbf{\nabla} h - f \hat{\mathbf{n}} \times \mathbf{u}.\label{eq:momentum}
\end{eqnarray}
The Coriolis parameter $f = 2\boldsymbol{\Omega} \cdot \hat{\mathbf{n}}$ is twice the projection of the planetary angular rotation vector on the local vertical axis $\hat{\mathbf{n}}$ and $g$ is the constant of gravitational acceleration. 
When linearized about a state of rest ($\mathbf{u}=0$) and mean height ($h=H$), this dynamical system may be rewritten as 
$i \partial_t \Psi = \mathcal{H} \Psi$, where $\Psi = (\mathbf{u},\eta)$ is a triplet of fields describing the two components of the perturbed velocity field and the perturbed height field $\eta$, and where  $\mathcal{H}$ is a Hermitian operator \cite{supp}.  Because the fields $(\mathbf{u},\eta)$ are real, the operator $\mathcal{H}$ is equal to the negative of its complex conjugate: $\Xi \mathcal{H} \Xi^{-1} = -\mathcal{H}$ where $\Xi$ is the operator that effects complex conjugation, with $\Xi^2=1$. In the quantum context, the operation is referred to as a particle-hole transformation because it inverts the spectrum. Time reversal symmetry  $t\rightarrow -t$, $\mathbf{x}'\rightarrow \mathbf{x}$, $\eta\rightarrow \eta$, $\mathbf{u}\rightarrow -\mathbf{u}$ is broken by non-zero Coriolis parameter $f \ne 0$ in Eq. (\ref{eq:momentum}).  The broken symmetry generates  gaps in the shallow water spectrum \cite{vallis2017atmospheric}.

\begin{figure}
\center
\includegraphics[height=7cm]{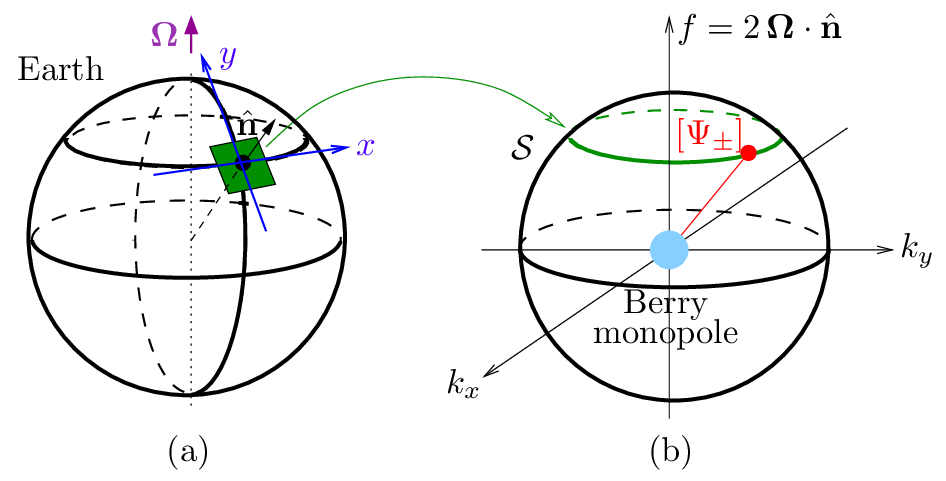} 
\caption{(a) Relation between the spherical geometry of a rotating planet and the unbounded $f$-plane geometry: At a given latitude, the flow is assumed to take place in the tangent plane, and the Coriolis parameter $f$ is twice the vertical component of the Earth's rotation. (b) Parameter space $(k_x,k_y,f)$ for the eigenmodes on the unbounded $f$-plane geometry. The wave bands $\omega_+$, $\omega_-$, $\omega_0$ are well defined everywhere except at the origin which is a band-crossing point. We show that the set of eigenmodes $\Psi_{\pm}$ parameterized on any closed surface (here a sphere) enclosing this band crossing point posses singularities that are quantified by a Chern number.  This  is an integer that can be computed by integrating over this surface a local Berry curvature that depends on the eigenmodes. The curvature can be viewed as generated by a Berry monopole located at the band-crossing point.}
\label{figure2}
\end{figure}

The $f$-plane approximation commonly used in geophysics~\cite{vallis2017atmospheric} amounts to the neglect of Earth sphericity by assuming that the dynamics take place on a tangent plane with constant $f$ (Figure 2a). Translational symmetry ensures that eigenmodes of the linearized dynamics in this geometry are of the form $\hat{\Psi} \exp\left( i \omega t -i k_x x - i k_y y\right)$ where $\hat{\Psi} $ has three components. Viewing $f/c$ as an external parameter, where $c = \sqrt{gH}$ is the speed of non-rotating shallow water gravity waves, the eigenmodes may be easily found at each point in the space $(k_x, k_y, f/c)$ as depicted in Figure 2b.  There are 3 bands with frequencies $\omega_{\pm} = \pm\left(f^2+c^2 k^2 \right)^{1/2}$ and $\omega_0=0$ where $k^2 \equiv k_x^2 + k_y^2$, with corresponding wavefunctions $\{\Psi_{\pm},~ \Psi_0\}$. For $f \ne 0$, the bands separate by gaps of frequency $f$ (Figure 3).  The zero-frequency modes are in geostrophic balance; the other two modes are Poincar\'e waves with dispersions $\omega_{\pm}$ that are symmetric with respect to the origin in ($k_x,k_y,\omega$) space.  

\begin{figure}
\center
\includegraphics[height=9cm]{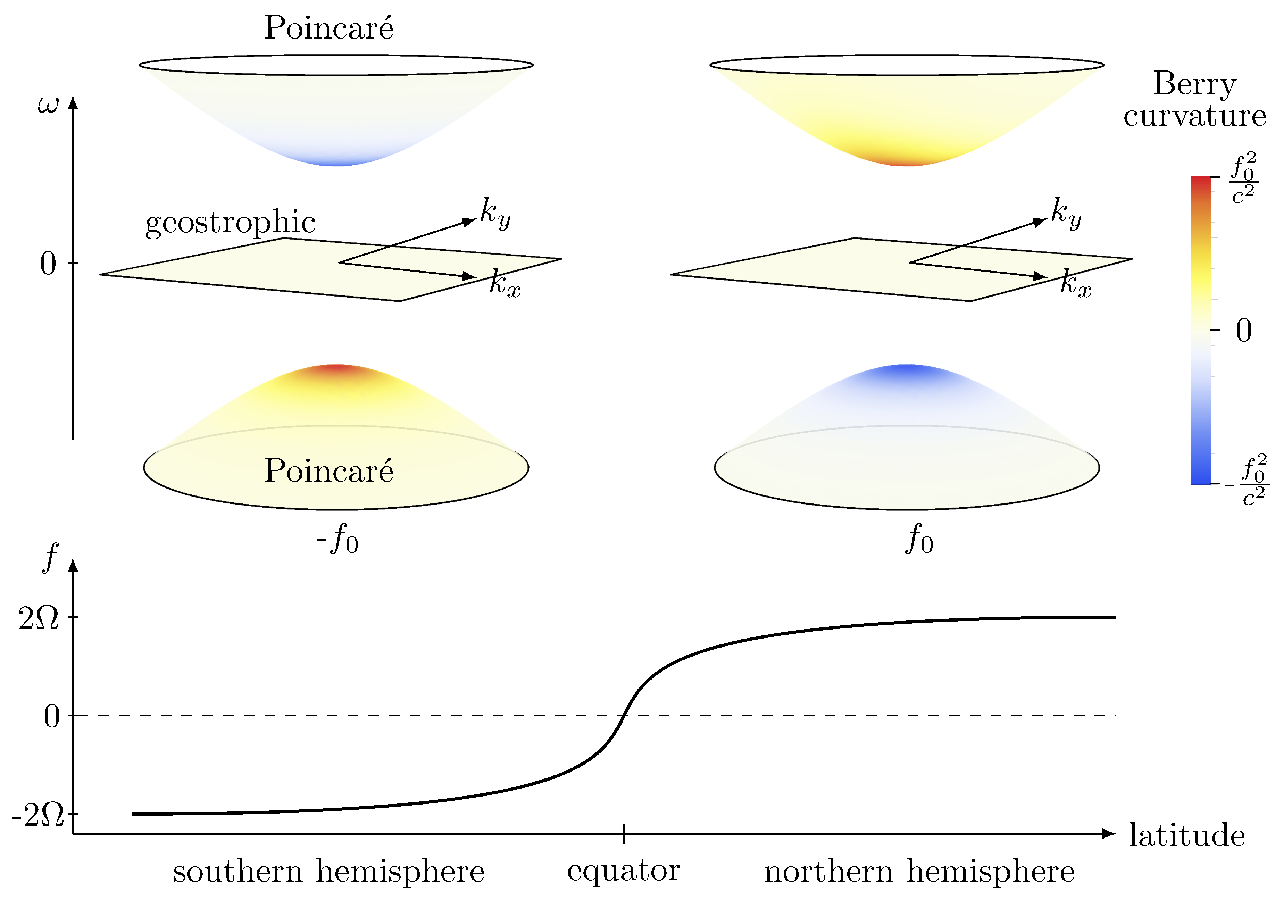}
\caption{Dispersion relation in unbounded $f$-plane geometry for the two signs of $f$. The color indicates the Berry curvature $B_n \equiv -\ii \nabla_p \times \left(\Psi_n^{\dagger} \nabla_p \Psi_n\right)$ for each wave band indexed by $n\in \{-,0,+\}$. The Berry curvature of the Poincar\'e bands is $B_\pm = \pm fc^2 (f^2+c^2(k_x^2+k_y^2))^{-3/2}$. It is concentrated around $k=0$, with extremal value $\pm c^2/f^2$, and switches sign as $f$ changes sign. The curvature vanishes for the geostrophic band. When integrated over the whole plane $(k_x,k_y)$, the Berry fluxes in the three bands give integers $(-1,0,1)$ for $f>0$ and $(1,0,-1)$ for $f<0$, consistent with the triplet of Chern numbers $\{ \Delta \mathcal{C}_-, \Delta \mathcal{C}_0, \Delta \mathcal{C}_+ \} =\{-2,0,2\}$. This shows that the set of delocalised bulk Poincar\'e modes cannot be continuously deformed from one hemisphere to another.}
\label{dispersion_berry}
\end{figure}

Eigenmodes depend on the triplet of parameters $(k_x, k_y, f/c)$ that correspond to the set of waves in all possible $f$-plane configurations. The eigenmodes do not vary with the distance from the origin in $(k_x, k_y, f/c)$-space and can therefore be  parameterized on the surface of a sphere $\mathcal{S}$ that encloses the singular band-crossing point at the origin $(k_x, k_y, f/c)=(0, 0, 0)$ [see Figure 2b and \cite{supp}].  Each of the eigenstates $\left\{ \Psi_-, \Psi_0, \Psi_+ \right \}$ defines a fiber bundle over $\mathcal{S}$ that may  possess topological defects. The singularities reflect the impossibility of continuously defining the eigenmodes everywhere on the sphere, and in particular over both of Earth's two hemispheres simultaneously. They are quantified by the first Chern number { $\Delta \mathcal{C}$ that can be calculated for each bulk mode $n$ as the flux of the Berry curvature $B_n= -\ii \nabla_p \times \left(\Psi_n^{\dagger} \nabla_p \Psi_n\right)$ through the sphere $\mathcal{S}$, with $\Psi^{\dagger}_n$ the conjugate transpose of $\Psi_n$, and $\nabla_p=\left(\partial_{k_{x}},\partial_{k_{y}},\partial_{f/c}\right)$. In other words, there exists a }quantized Berry flux generated by a (Berry) monopole located at the center of $\mathcal{S}$, where the three bands cross \cite{berry1984quantal,bernevig2013topological}. The singularities are analogous to the one exhibited by an electron wavefunction that cannot be defined continuously around a Dirac magnetic monopole \cite{dirac1931quantised}. We find $\{ \Delta \mathcal{C}_-, \Delta \mathcal{C}_0, \Delta \mathcal{C}_+ \} =\{-2,0,2\}$ \cite{supp}, namely only the Poincar\'e modes $\Psi_{\pm}$ are topologically non-trivial as the geostrophic modes $\Psi_0$ have zero Chern index, in agreement with the bulk-boundary correspondence \cite{hatsugai1993chern,fukui}. 

To understand qualitatively the correspondence between these bulk properties and the emergence of unidirectional edge states in the presence of an equator, is it worth considering the case of a planar flow in an unbounded domain with  $f$ varying in the $y$-direction from $-2\Omega$  to $2\Omega$ (see Figure 3). Far from the interface, the eigenmodes are given by delocalized solutions, i.e. by those computed in the case of constant $f$.  If one could continuously deform the whole set of positive frequency eigenmodes from one hemisphere to other, for instance by varying $f$ slowly with y, then the eigenmodes would be given by solutions close to those calculated for constant $f$. Our previous calculation shows that this continuous deformation is prohibited by the occurance of $\Delta C_+=2$ phase singularities (positive vortices) when the plane $f=0$ is crossed. In order to remove these two singularities, the positive frequency band and the negative frequency bands must be connected to each other as the sum of their Chern numbers is zero. This connection happens through the emergence of two edge states that fill the frequency gaps.  For any frequency that lies within the bulk gaps, the number of topological edge states is fixed by the set of Chern numbers~\cite{hatsugai1993chern}. Because $\Delta \mathcal{C}_{\pm} = \pm 2$ there are two extra unidirectional edge modes in the frequency gaps \cite{supp}. 
 
It is instructive to examine the Berry curvature for the Poincar\'e modes. As shown in Figure 3, the curvature is mainly concentrated around $k=0$ where it reaches extremal values, and importantly, changes sign with $f$. It follows that its flux for each Poincar\'e mode $\mathcal{C}_\pm=\frac{1}{2\pi}\int_{-\infty}^{\infty} \rm{d}k_x \rm{d}k_y B_\pm = \pm \text{sgn}(f)$ is an integer that only depends on the hemisphere. 
It is thus tempting to say that the Poincar\'e eigenmodes on the two hemispheres are topologically distinct by interpreting $\mathcal{C}_\pm$ also as a Chern number, as the difference $\mathcal{C}_\pm(f>0)-\mathcal{C}_\pm(f<0) =\pm2$ coincides with the first Chern number $\Delta \mathcal{C}_\pm$. This would be rigorously true if the two-dimensional manifold through which this Berry flux is computed at fixed $f$ were closed, for instance when the wave-numbers $(k_x,k_y)$ live on a Brillouin zone that reflects an underlying lattice. For continuous fluids, only $\Delta \mathcal{C}_\pm$ is a well defined topological number, but this suffices to characterize the topological property of the bulk modes, and thus the existence of the two equatorial unidirectional modes.

We stress one important point concerning the role of the spherical geometry of the planet in our approach. We removed this sphericity with the $f$-plane approximation, equivalent to holding the Coriolis parameter constant in space.  However, through the construction of the sphere $\mathcal{S}$ in parameter space $(k_x, k_y, f/c)$, we recover the effect of a varying Coriolis parameter $f$ on the shallow water eigenmodes.  In this way, sphericity works its way back into the problem. The detailed geometry of the Earth is no longer needed as topology itself requires the existence of Yanai and Kelvin waves.  Even a misshapen sphere would support the waves. 

Topology guarantees the existence of equatorial Yanai and Kelvin waves, obviating the need to carry out the classic but more complex calculation on the equatorial beta plane \cite{vallis2017atmospheric}.  On the equatorial beta plane, Rossby and Poincar\'e waves can also be equatorially trapped.  However, this trapping depends on the precise longitudinal variation of $f(y)$, as may be demonstrated numerically.   In contrast, the topological origin of Kelvin and Yanai modes makes them insensitive to the details of the interface, such as the detailed shape of $f(y)$ \cite{supp}.  We also performed numerical scattering experiment showing that there is no possibility for Kelvin or Yanai wave excited within the bulk frequency gap, away from the other bands, to exchange energy with other modes that propagate energy westward \cite{supp}. Consequently, there is no energy backscattering in the presence of topography. This robustness against disorder can now be understood as a consequence of topology. 

Other ideas from topology have been applied to hydrodynamics~\cite{arnold1999topological,moffatt2001topology,kleckner2013creation}.  However, the appearance  of singularities in the set of eigenmodes that arises from the breaking time-reversal symmetry has so far been overlooked in this context, as well as the striking physical consequence of unidirectional edge modes filling the frequency gaps.  The general principle of bulk-boundary correspondence may now be applied to other fluid systems of interest.

The shallow water system exhibits particle-hole symmetry stemming real-valued velocity and displacement fields.  More generally, any linearized fluid flow model that can be written in terms of an hermitian operator that breaks time reversal symmetry belong to the symmetry class with Cartan label D, which means that non trivial topological properties may arise \cite{kitaev2009periodic,ryu2010topological}. Other physical systems that may belong to class D are chiral p-wave superconductors~\cite{bernevig2013topological,kallin2012} and superfluid $^3$He-A~\cite{Ikegami2013}.  The linear operator of flow dynamics can be non-Hermitian in the presence of mean-flows and dissipation, in which case other topological properties may appear \cite{Zeuner:2015fd}. We expect that topology may ultimately shape the global structure of a number of other astrophysical and geophysical wave spectra, where similar gaps opened in the presence of symmetry breaking fields are known to exist. For instance, Lamb waves are edge states that fill the gap between acoustic and gravity waves because gravity breaks another discrete symmetry, that of inversion.   Hall magnetohydrodynamics is another possible setting for topological edge waves \cite{Witalis1986}.  It will also be also interesting to study in more detail the resilience of topological waves against dissipation, and non-linear wave-wave scattering processes.




\bibliographystyle{Science}

\section*{Acknowledgments}
We thank David Carpentier, Baylor Fox-Kemper, Thibaud Louvet, Jim Sauls, and Steve Tobias for discussions.   P.D. was supported by the French Agence Nationale de la Recherche (ANR) under Grant TopoDyn (ANR-14-ACHN-0031).

\section*{Supplementary materials}
1- First Chern Number of Shallow Water Eigenmodes\\
2- Numerical Calculation of Eigenmodes With $f(y)$: From the Shallow Water Model to a Haldane-like Model on the Lieb lattice\\
3 -Scattering of Equatorial Kelvin Waves by a Static Perturbation\\
Figs. S1 to S4\\
References \textit{(28-33)}


\pagebreak

\textwidth = 6.5 in
\textheight = 9 in
\oddsidemargin = 0.0 in
\evensidemargin = 0.0 in
\topmargin = 0.0 in
\headheight = 0.0 in
\headsep = 0.0 in
\parskip = 0.2in
\parindent = 0.0in

\begin{center}
\textbf{\large Supplementary Material for ``Topological Origin of Equatorial Waves''}
\end{center}
\setcounter{equation}{0}
\setcounter{figure}{0}
\setcounter{table}{0}
\setcounter{page}{1}
\makeatletter
\renewcommand{\theequation}{S\arabic{equation}}
\renewcommand{\thefigure}{S\arabic{figure}}

\section{First Chern Number of Shallow Water Eigenmodes}

In the classical dispersion of equatorial waves on a beta plane ($f=\beta y$, Figure 1a in the Main Text), one recognizes the three bands obtained within the $f$-plane approximation, except that part of the degeneracy of geostrophic modes  is lifted, with the emergence of Planetary (Rossby) waves due to the spatial variation of the Coriolis parameter. In addition, when varying the zonal (directed along the equator) wavenumber $k_x$ from $-\infty$ to $+\infty$, the Poincar\'e band of positive (negative) frequency acquires (is depleted of) two states whereas the total number of states acquired by the Rossby band vanishes. Thus, the occurence of one-way equatorial waves in the dispersion relation can be seen as a \textit{spectral flow} from one band to another that inherits $\Delta \nu$ states, with $\Delta \nu_\pm = \pm 2$ and $\Delta \nu_0=0$. According to the bulk-boundary correspondence \cite{hatsugai1993chern,fukui}, such a spectral flow suggests a topological property of the bulk (Poincar\'e and Rossby) waves, obtained by assuming that the Coriolis , that encodes the impossibility to smoothly define each of these waves for any $k_x$, $k_y$ when the  $f$ can change sign. This correspondence lies on the index theorem that relates the index of the bulk Hamiltonian both to the spectral flow and to a topological index \cite{fukui}. More precisely, we expect that the spectral flow to the band $n$ is given by $\Delta\nu_n=\Delta \mathcal{C}_n$ where $\mathcal{C}_n$ is a topological invariant called the first Chern number assigned to the band $n$ \cite{hatsugai1993chern,fukui}. In the following, we compute the Chern numbers of the Poincar\'e and Rossby modes and show that their value indeed satisfy the bulk-boundary correspondence, thus unveiling the topological origin of the Kelvin and Yanai equatorial waves.

We consider flows on a $f$-plane with $x$ in the zonal direction (positive towards the East) and $y$ the meridional direction (positive towards to North) [see Figure 1(a) in the Main Text]. The velocity field is denoted by  $\mathbf{u}(\mathbf{x},t)=(u(\mathbf{x},t),v(\mathbf{x},t))$, with $\mathbf{x}=(x,y)$.  The value of the Coriolis parameter $f$  is spatially constant, but will be considered as an external parameter that can be varied between $\pm 2\Omega$, where $\Omega$ is the planet rotation rate. The extremal values  correspond to the value of $f$ at the Northern and  Southern pole. The system possess one intrinsic time scale $1/2\Omega$ and one intrinsic length scale given by the global Rossby radius of deformation $L_d$ along with a speed $c$:
\begin{equation}
L_d=\frac{c}{2\Omega},\quad c=\sqrt{gH}.
\end{equation}
The dynamics of Equations (1)-(2) can then be expressed in terms of dimensionless quantities 
\begin{equation}
 \tilde{\eta}=\frac{h-H}{H},\quad \tilde{\mathbf{u}}=\frac{\mathbf{u}}{c},\quad \tilde{t}= 2 \Omega t,\quad \tilde{f}=\frac{f}{2\Omega},\quad \tilde{\mathbf{x}}=\frac{{\mathbf{x}}}{L_d},
\end{equation}
and the linearized equations of motion around a state at rest ($\mathbf{u}=0$, $h=H$) read
\begin{eqnarray}
\partial_{\tilde{t}} \tilde{\eta} &=& -\partial_{\tilde{x}} \tilde{u} -\partial_{\tilde{y}} \tilde{v} \label{eq:lin1}\\
\partial_{\tilde{t}}  \tilde{u} &=&-\partial_{\tilde{x}} \tilde{\eta} +\tilde{f} \tilde{v} , \label{eq:lin2}\\%
\partial_{\tilde{t}}  \tilde{v} &=&-\partial_{\tilde{y}} \tilde{\eta} -\tilde{f} \tilde{u} \label{eq:lin3}  \, .
\end{eqnarray}
For convenience, we drop the $\tilde{}$ tilde in the following. 

Eigenmodes (equivalently, normal modes) are propagating planar waves. Let $\hat{\eta}$, $\hat{u}$, $\hat{v}$ be the amplitudes of the fields $ {\eta}$, $ {u}$, $ {v}$ for mode $e^{i {\omega}  {t}  -ik_x x-i k_y y}$. The amplitudes obey the eigenvalue equation:
\begin{equation}
\omega \begin{pmatrix} \hat{\eta} \\ \hat{u} \\ \hat{v} \end{pmatrix}=  \begin{pmatrix} 0 & k_x & k_y \\  k_x  &0 & -i {f} \\  k_y  & i  {f} & 0\end{pmatrix} \begin{pmatrix} \hat{\eta} \\ \hat{u} \\ \hat{v} \end{pmatrix} \, .\label{eq:EigProb}
\end{equation}
Diagonalization of this matrix leads to three eigenvalues corresponding to three wave bands in wavenumber-frequency space. The normalized eigenvectors associated with the positive frequency band $ {\omega}_{+} =\left(  k^2 +  {f}^2\right)^{1/2}$ are
\begin{equation}
\Psi_+(k_x,k_y, {f})=\frac{1}{\sqrt{2}}\begin{pmatrix} \frac{ k}{\sqrt{k^2 +  {f}^2}}\\ { \frac{k_x}{k}-i\frac{ {f}k_y}{k\sqrt{k^2 +  {f}^2}}} \\ { \frac{k_y}{k}+i\frac{fk_x}{k\sqrt{k^2 +  {f}^2}}}\end{pmatrix}
\end{equation}
The ones associated with the zero frequency band $\omega_0=0$ are
\begin{equation}
\Psi_0(k_x,k_y, {f})=\frac{1}{\sqrt{ k^2 + {f}^2}}\begin{pmatrix}  f \\ i  {k_y} \\ { -i {k_x}}\end{pmatrix}
\end{equation}
Eigenvectors for the negative frequency band  $ {\omega}_{-} =-\left(  k^2 +  {f}^2\right)^{1/2}$ are easily obtained by taking advantage of the system symmetry: $\Psi_-(k_x,k_y, {f})=\Psi_+(-k_x,-k_y,- {f})$.\\ 

We calculate the Chern number of the fiber bundle defined by eigenvectors of the positive frequency wave band parameterized on the unit sphere in the space $\left(k_x,k_y, {f}\right)$ [see Figure 1(b) in the Main Text]. Importantly, normalized eigenvectors $\Psi_0$ and $\Psi_{\pm}$ are defined only up to a phase.   Here we show that this phase cannot be defined continuously for $\Psi_{\pm}$ over all possible values of $\left(k_x,k_y, {f}\right)$.   To see this, first notice that the eigenstates can be parametrized  with the use of spherical coordinates.  Introducing the colatitude $\theta$ and the longitude $\varphi$ (that should not be confused with colatitude and longitude on the planetary sphere in physical space), the normalized eigenvectors of the positive frequency wave band may be rewritten as:
\begin{equation}
\Psi_+=\frac{1}{\sqrt{2}}\begin{pmatrix} \sin \theta \\ \cos \varphi -i \cos\theta \sin \varphi   \\\sin\varphi +i\cos \theta \cos \varphi\end{pmatrix} \ .
\end{equation} 
The eigenvectors are multivalued at $\theta=0$  and $\theta = \pi$, respectively the north and south poles of the parameter sphere $\mathcal{S}$  shown in Fig. 1(b) of the Main Text.  The multivaluedness is a topological property of the vector bundle that is defined as the set of $\Psi_{+}$ over the base space $\left(\theta, \varphi\right)$.  The multivaluedness can be quantified by non-vanishing first Chern number that we evaluate below.
  
Recalling that the eigenmodes are defined only up to a phase, a gauge transformation can be used to regularize the eigenmodes at one of the poles. 
In particular, we define
\begin{equation}
\Psi_+^N=\Psi_+e^{i\varphi},\quad \Psi_+^S=\Psi_+e^{-i\varphi}
\end{equation}
so that $\Psi_+^N$ is a single-valued function of $(\theta, \varphi)$ everywhere but at the south pole.  Likewise $\Psi_+^S$ is a single-valued function of $(\theta, \varphi)$ everywhere but at the north pole.  The Berry connections associated with these vector bundles are
\begin{equation}
\mathbf{A}_+^N=-i\left<\Psi_+^N ,\nabla_s \Psi_+^N\right>,\quad \mathbf{A}_+^S=-i\left<\Psi_+^S , \nabla_s \Psi_+^S\right>
\end{equation}
with the inner product defined for the 3-component amplitudes as $\left<A,B\right> \equiv A_1^*B_1+A^*_2B_2+A_3^*B_3$ and $\nabla_s$ is the gradient operator on the unit sphere in $(k_x,k_y,f)$ parameter space.
These Berry connections are related by the gauge transformation through 
\begin{equation}
\mathbf{A}_+^S=\mathbf{A}_+^N-2\nabla_s \varphi \label{eq:A+_A-} \, .
\end{equation}
At any point where $\mathbf{A}_+^N$ and $\mathbf{A}_+^S$ are not singular, the Berry curvature is given by 
\begin{equation}
\mathbf{B}_+=\nabla_s\times \mathbf{A}_+^N=\nabla_s\times \mathbf{A}_+^S \, .
\end{equation}
The Chern number for the fiber bundle of eigenvectors with positive frequencies on the unit sphere in $(k_x,k_y,f)$-space is defined as the flux of the Berry curvature through $\mathcal{S}$
\begin{equation}
\Delta \mathcal{C}_+=\frac{1}{2\pi}\int_0^{\pi} \mathrm{d} \theta \int_0^{2\pi} \mathrm{d} \varphi\  \mathbf{B}_+\cdot \mathbf{e}_k 
\end{equation}
where $\mathbf{e}_k(\theta,\varphi)$ is the vector normal to the sphere pointing in the outward direction. Stokes theorem can be applied separately for each hemisphere of $\mathcal{S}$ where the Berry connection is well defined, e.g. along the equator circle ($\theta=\pi/2$), with  $\mathbf{B}_+=\nabla_s\times \mathbf{A}_+^N$ in the northern hemisphere and  $\mathbf{B}_+=\nabla_s\times \mathbf{A}_+^S$  in the southern hemisphere.  The calculation yields
\begin{equation}
\Delta \mathcal{C}_+=\frac{1}{2\pi} \int_0^{2\pi} \mathrm{d} \varphi \left(\mathbf{A}_+^N-\mathbf{A}_+^{S}\right) \cdot \mathbf{e}_\varphi.
\end{equation}
Using Eq. (\ref{eq:A+_A-}), we find $\Delta \mathcal{C}_+=2$. This Chern number can be interpreted as the topological invariant of the transition from one hemisphere to another for the case of the positive frequency eigenmodes of the shallow water model in the $f$-plane approximation, with $f$ taken as an external parameter that tunes this transition.  Note that we have derived this result by considering the fiber bundle of eigenmodes parameterized on the unit sphere, but it is straightforward to show that it holds on any sphere of radius $\omega_+>0$ in $(k_x,k_y,f)$ space since the normalized eigenvectors $\Psi_+$ do not depend on $\omega_+$. 
Similar calculations show that the Chern number for the negative frequency eigenmodes is $\Delta \mathcal{C}_-=-2$, and $\Delta \mathcal{C}_0=0$ for the zero-frequency eigenmodes.

According to bulk-edge correspondence~\cite{hatsugai1993chern,bernevig2013topological}, the number of topological edge states at the interface between two domains of opposite $f$ is fixed by the set of Chern numbers $\left\{\Delta \mathcal{C}_-,\ \Delta \mathcal{C}_0,\ \Delta \mathcal{C}_+\right\}$. Let $N^{\text{above}}_{\text{right}}-N^{\text{above}}_{\text{left}}$ be the difference in the number of right-moving and left-moving modes in the gap above the central geostrophic band, and $N^{\text{below}}_{\text{right}}-N^{\text{below}}_{\text{left}}$ be the difference in the number of right-moving and left-moving modes in the gap below.  Bulk-edge correspondence then states that $|\Delta \mathcal{C}|=\left| \left(N^{\text{above}}_{\text{right}}-N^{\text{above}}_{\text{left}}\right)-\left(N^{\text{below}}_{\text{right}}-N^{\text{below}}_{\text{left}}\right)\right|$.  The numbers can vary with frequency within the gap, but the difference 
$|\Delta \mathcal{C}|$ is the number of topologically protected chiral edge modes. In the present case, Chern number $\Delta \mathcal{C}_0=0$ means that there are equal numbers of topological edge states with the same chirality  in the gaps above and below the central band of geostrophic modes. Since there is no gap above the positive frequency band, and no gap below the negative frequency band, the number of chiral edge states in each gap surrounding the central band is given by $\left|\Delta \mathcal{C}_{\pm}\right|=2$.

\section{Numerical Calculation of Eigenmodes With $f(y)$: From the Shallow Water Model to a Haldane-like Model on the Lieb lattice}

The linearized rotating shallow water model on a doubly periodic domain with $f$ varying in the $y$ direction is diagonalized numerically using a standard C-grid discretization procedure~\cite{cushman2011}. This approach establishes a direct link to lattice models used in condensed matter physics. 

The domain is discretized into a uniform grid with $N_x \times 2 N_y$ unit cells located at  $(x,y) = (m\, \delta x, n\, \delta y)$ where $\delta x$ and $\delta y$ are the size of the grid cells, and $m$ and $n$ are integers.  Assuming time-harmonic solutions, eigenmodes of the discretized shallow water system are described at each grid point by the real part of the triplet $e^{\ii\omega t} (h_{m,n},\mathbf{u}_{m,n})$. Following the C-grid discretization procedure, the system of equations (\ref{eq:lin1})-(\ref{eq:lin2})-(\ref{eq:lin3}) then turns into an eigenvalue problem for the fields $(h(m,n,\omega), \mathbf{u}(m,n,\omega))$:
\begin{subequations}
\begin{align}
\omega\, u_{m,n} &= -t_x \left( h_{m-1,n} - h_{m,n} \right) - \frac{\ii f}{4}\left( v_{m,n} + v_{m,n+1} + v_{m-1,n} + v_{m-1,n+1} \right)\\
\omega\, v_{m,n} &= -t_y \left( h_{m,n-1} - h_{m,n} \right) +\frac{\ii f}{4} \left( u_{m,n} + u_{m+1,n} + u_{m+1,n-1} + u_{m,n-1} \right)\\
\omega\, h_{m,n} &= t_x \left( u_{m,n}-u_{m+1,n} \right) +t_y \left( v_{m,n} - v_{m,n+1}\right)
\end{align}
\end{subequations}
where the change of variable $\ii h \rightarrow h$ was performed for convenience, and where we have defined $t_x=\frac{1}{\delta x}$ and $t_y=\frac{1}{\delta y}$.  
We consider a meridional dependence of the Coriolis parameter through the map $f_y$ that we specify below. This allows us to treat arbitrary variations of $f$, and in particular any equator-like interface from $f<0$ to $f>0$. For concreteness, we consider periodic boundary conditions in the $y$ direction with $n$ running from $1$ to $ 2 N_y$, and we choose 
\begin{equation}
f_n=\tanh{\left((n-1)\frac{\delta y}{R_f} \right)} -\tanh{\left((n-N_y)\frac{\delta y}{R_f}\right)} + \tanh{\left((n-2N_y)\frac{\delta y}{R_f}\right)}
\end{equation}
where $R_f$ gives the scale of variation for the gradient of the Coriolis parameter in the $y$ direction. There are two regions with opposite $f$ and two interfaces between the regions (see Figure \ref{fig:f}). The doubly periodic geometry has the advantage of simplifying numerical computations. The price to pay is the existence of tow equators. \\
\begin{figure}[h]
\center
\includegraphics[scale=1]{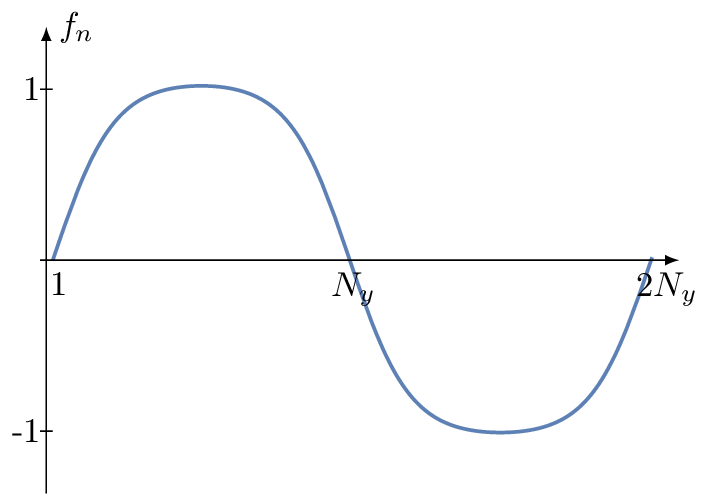} 
\caption{Variation of the Coriolis parameter as a function of $n$ for $N_y=50$ and $R_f/\delta y= 10$.}
\label{fig:f}
\end{figure}

The frequency $\omega$ {is the eigenvalue} of a tight-binding model on the Lieb lattice shown in Figure \ref{fig:lieb}.
Each site is coupled to its nearest neighbors by hopping parameters $\pm t_x$ and $\pm t_y$. The minus sign that appears in these matrix elements can be interpreted as a phase of $\pi$ picked up when hopping from site to site. While this phase does not break time-reversal symmetry, the coupling $\pm \ii f/4$ that accounts for a phase of $\pm \pi/2$ accumulated between second nearest neighbors, does break it.  The total phase accumulated around a unit cell is a multiple of $2\pi$, so that the net flux through the system vanishes. This behavior is analogous to the seminal Haldane model on the honeycomb lattice, the archetypal example of a topological Chern insulator in the absence of an applied magnetic field \cite{haldane88}.\\

\begin{figure}[h]
\center
\includegraphics[scale=1]{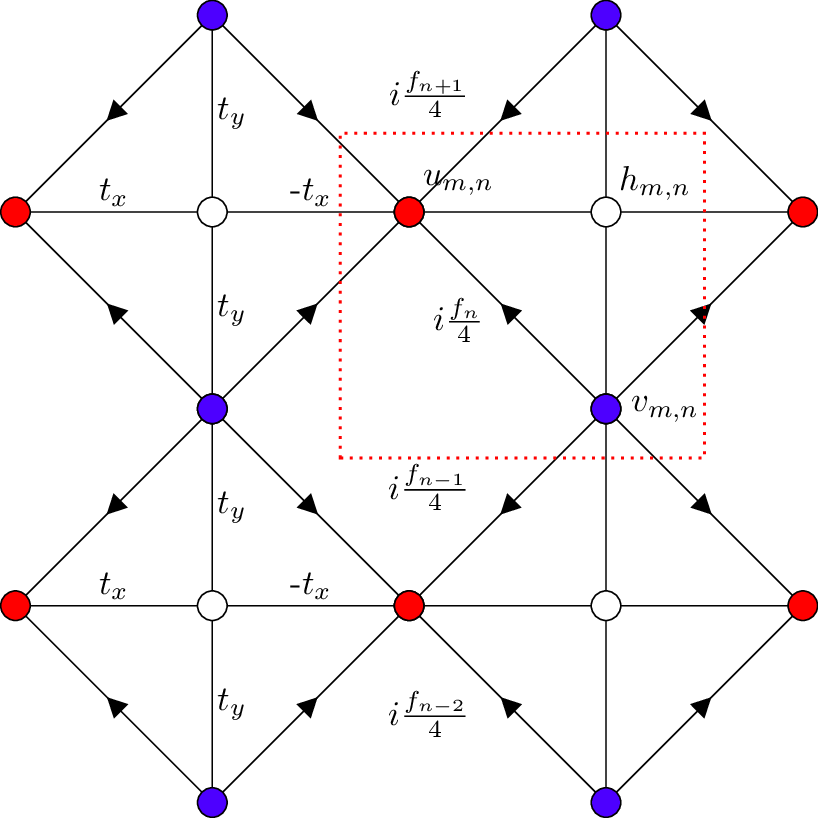} 
\caption{Lieb lattice representing the coupling between the three fields of the discretized rotating shallow water model following the C-grid procedure. The field $u(m,n)$ is represented by a red dot, $v(m,n)$ by a blue dot and $h(m,n)$ by a white dot. A unit cell is highlighted by a red dashed rectangle. The nearest neighbors are coupled by $\pm t_x$ and $\pm t_y$ hopping terms, and the second nearest neighbors by an imaginary terms proportional to the Coriolis parameter $f$ that breaks time-reversal symmetry.  Arrows indicate directions along which phase $+\pi/2$ is picked up.}
\label{fig:lieb}
\end{figure}

Because of translational invariance, the Fourier transformed fields $(h(k_x,n,\omega), \mathbf{u}(k_x,n,\omega))$ satisfy
\begin{subequations}
\begin{align}
\omega\, u_{k_x,n} &= -t_x \left( \ee^{-\ii k_x \delta x} -1 \right) h_{k_x,n} - \ii  \frac{f_n}{4}\left(1+\ee^{-\ii k_x \delta_x} \right) \left(v_{k_x,n} + v_{k_x,n+1} \right) \\
\omega\, v_{k_x,n} &= -t_y\left(  h_{k_x,n-1} - h_{k_x,n}  \right)+ \ii  \frac{f_n}{4}\left(1+\ee^{\ii k_x \delta_x} \right) \left(u_{k_x,n-1} + u_{k_x,n} \right)\\
\omega\, h_{k_x,n} &= t_x \left(1-\ee^{\ii k_x \delta_x} \right) u_{k_x,n}+t_y \left( v_{k_x,n} - v_{k_x,n+1}\right) \ .
\end{align}
\label{eq:syst}
\end{subequations}

There are four dimensionless parameters given by $R_f/L_y$ $L_d/L_y$, $L_x/L_y$, and $\delta y/L_y$. The first parameter controls the interface thickness between regions of constant (but opposite) $f$. It is of order $1$ for a spherical geometry, but the difference between non-topological and topological edge modes is enhanced by considering a sharper interface, with $R_f/L_y\approx 0.1$.  The second non-dimensional  parameter, $L_d/L_y$, controls the intrinsic length scale of the domain in comparison to the domain size. It is of order one in the atmosphere, and of order $0.1$ in the ocean.  The third parameter $L_x/L_y$ is a geometric parameter, of order one for the atmosphere and oceans on Earth. For convenience, we consider  $L_x/L_y\gg 1$, which yields continuous lines $\omega(k_x)$ in the dispersion relation. The last parameter is related to the discretization of the continuous problem and is chosen such that  $\delta y/L_y\ll R_f/L_y$ and $\delta y/L_y\ll L_d/L_y$. The precise form of the dispersion relation for $k_x \sim \pm 2\pi / \delta_y$ depends on the choice of the discrete geometry.  However, for a given $k_x$, the dispersion relation of the continuous shallow water system is recovered in the limit $\delta x \rightarrow 0$.\\

The system (\ref{eq:syst}) is diagonalized numerically. The frequency spectrum shown in Fig. \ref{figureSuppTorus} is obtained for $N_y=50$, $R_f/\delta y =10$ and Rossby radius of deformation $L_d=R_f/2$.  We have checked that regardless of the value of these dimensionless parameters, the gaps obtained for $f\neq 0$ are filled by two equatorially trapped modes, and that for each equator, the difference in the number of right-moving and left-moving edge modes within the frequency gaps is equal to $\pm 2$: in the earth-like equator ($\partial_y f>0$), these unidirectional modes have eastward group velocity. By contrast, the amplitude of the Rossby waves, and the emergence of equatorially trapped Poincar\'e waves, depend on the shape of $f(y)$.

\begin{figure}[h!]
\center
\includegraphics[width=0.4\textwidth]{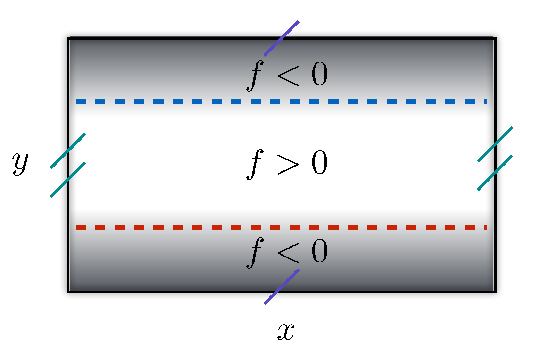} 
\hskip 2cm
\includegraphics[width=0.4\textwidth]{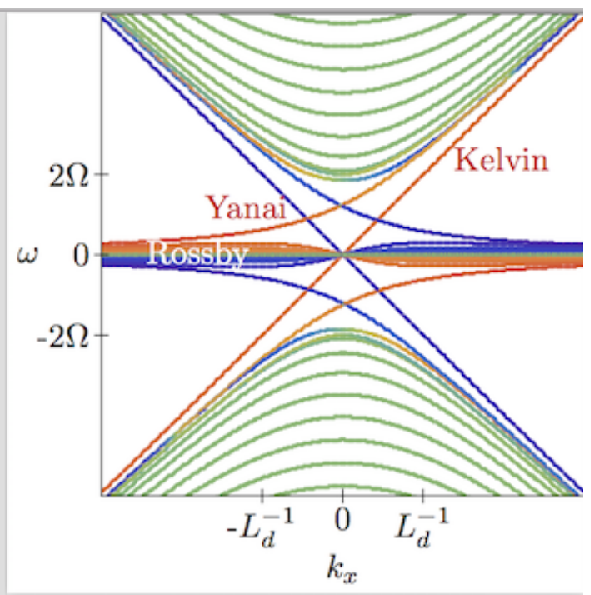}
\caption{(a) A plane that is periodic in both the x- and y-directions, with the Coriolis parameter $f$ varying between $\pm2\Omega$.  The region with $f=2\Omega$ (white) is separated from another with $f=-2\Omega$ by two narrow equatorial regions (blue and red dashed lines) over which $f(y)$ interpolates smoothly.  (b) Dispersion relation for modes obtained by numerical diagonalization of the linearized shallow water model on a lattice.  $L_d=c/ 2\Omega $ is the Rossby radius of deformation. The delocalized bulk modes are marked in green, edge states localized around the earth-like equator (with $\partial_y f> 0$ at the equator) are red, and edge states localized on the other equator are blue. Note the existence of a frequency gap filled by two unidirectional edge modes propagating along each equator.}
\label{figureSuppTorus}
\end{figure}

\section{Scattering of Equatorial Kelvin Waves by a Static Perturbation}

Equatorial Kelvin waves were predicted in Refs. \cite{Matsuno:1966wt} and subsequently observed in Earth's oceans~\cite{miller1988geosat} and atmosphere~\cite{wheeler1999convectively}.  The waves are reproduced by models more general than the single-layer shallow water equations considered in the Main Text \cite{gill1982atmosphere}.  Conversion of low-frequency eastward-moving Kelvin waves to westward Rossby waves can occur at the eastern margins of ocean basins \cite{cushman2011} or by scattering from topography \cite{McPhaden:1987ck,Majda:1999gt}.  Here we investigate the frequency dependence of the scattering of equatorial Kelvin waves by a region of enhanced stratification in a 2-layer primitive equation model
to demonstrate that Kelvin waves that oscillates at an intermediate frequency in the gap between the Rossby and Poincar\'e waves are not backscattered by the perturbation.  

The model is defined on a spherical geodesic grid as described in Ref. \cite{Marston:2012co}. 
This minimal model of the general circulation describes fluid motion in the horizontal within each layer, {flow between the layers}, and the variation of temperature within each layer.  The model has rigid upper and lower lids and supports internal waves at the interface between the two layers.  The temperature field is advected by the fluid and acts back upon it via pressure as determined by the equation of state for an ideal gas and hydrostatic balance.   There are in total 5 independent dynamical fields once the incompressibility of the fluid has been taken into account. The idealized model can be used to describe either a planetary atmosphere or an ocean that covers the whole planet.  Code that implements the model is publicly available\footnote{The application ``GCM'' is available for macOS 10.9 and higher on the Apple Mac App Store at URL http://appstore.com/mac/gcm}.

\begin{figure}[h!]
\center
\includegraphics[width=0.95\textwidth]{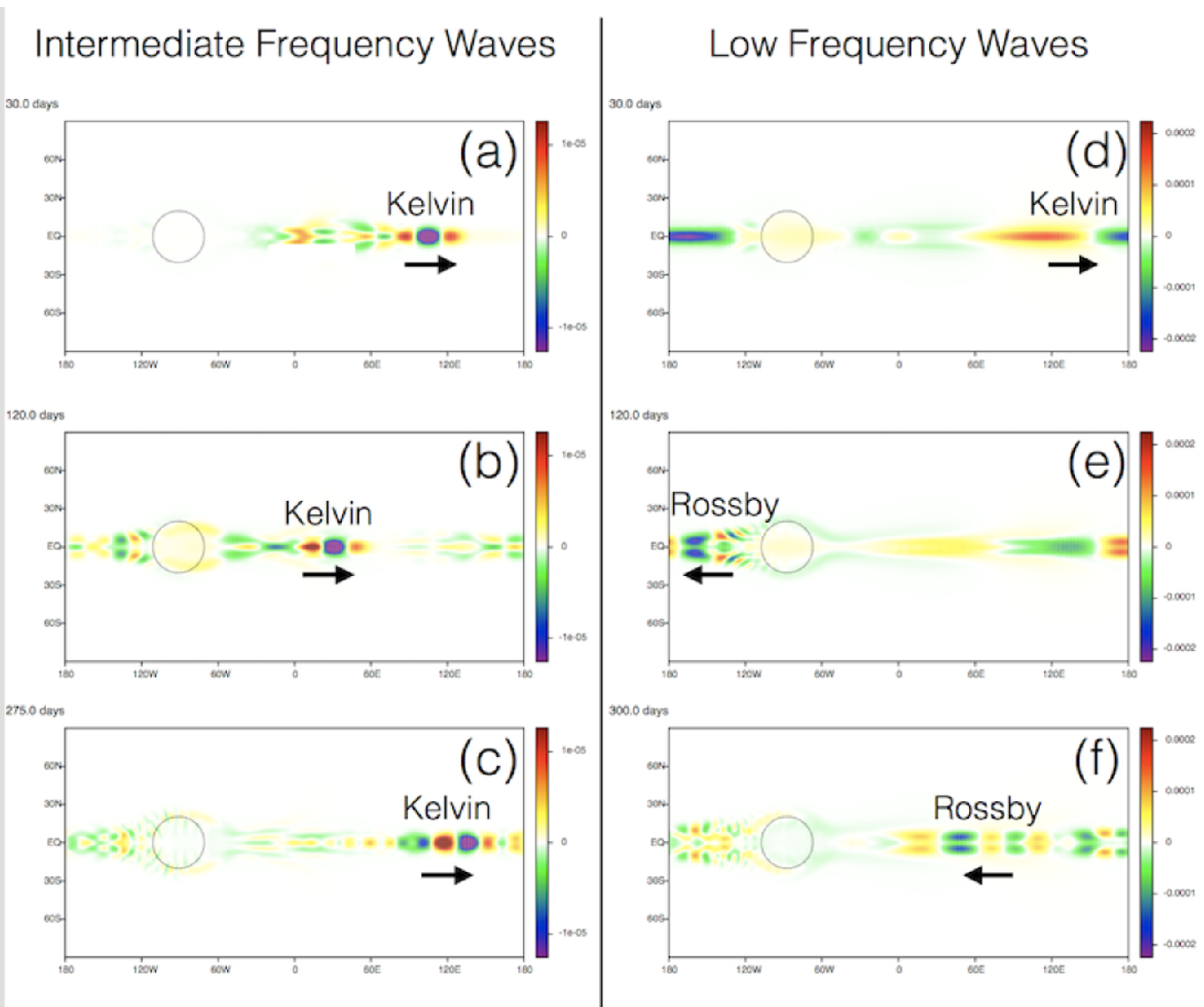} 
\caption{Simulated scattering of equatorial Kelvin waves.  The region of enhanced stability is indicated by the dashed circle.  The temperature of the upper layer in that region is enhanced by up to $10$K above an otherwise uniform value of $1$ K over that of the lower layer.  The color scale displays the vertical mean temperature.  (a) An intermediate frequency ($\omega \approx 1$ (day)$^{-1}$) eastward Kelvin wave packet at time $t = 30$ days prior to interaction.  The arrow indicates the direction of propagation.  (b) By $120$ days the intermediate frequency Kelvin wave has passed through the region of enhanced stability intact. (c)  After $275$ days the Kelvin wave has passed through the scattering region twice yet remains coherent.  (d) A low-frequency ($\omega \approx 0.2$ (day)$^{-1}$), long-wavelength Kelvin wave at time $t = 30$ days, prior to interaction, moves eastward.   (e) At $120$ days the low-frequency Kelvin wave strongly interacts with the region of enhanced stability, and generates westward moving Rossby waves.  The Rossby waves are bimodal across the equator \cite{cushman2011}.  (f) By $300$ days the Kelvin wave packet has been converted to Rossby waves that continue propagating westward. } 
\label{scattering}
\end{figure}

The vertical mean temperature is initially uniform over the sphere, and can be set to zero without loss of generality. Stable stratification is imposed by setting the temperature of the upper layer to be $1$ K greater than that of the lower layer, except in a region centered on the equator at longitude $90^\circ$ W (indicated by the dashed circles in Figure \ref{scattering}) where the temperature difference increases up to a maximum of $10$ K further with a Gaussian profile of width $20^\circ$.  The lower layer is thus denser than the upper layer and supports internal Kelvin and Rossby waves that move at speeds comparable to those observed in the equatorial Pacific ocean.  Waves can scatter off the region of enhanced stability because the wave speed is higher in that region.  Kelvin waves in the ocean can be observed from satellite via surface altimetry as water that warms also expands.  The imposed model stratification means that the simulated vertical mean temperature likewise serves as a proxy for the interface height.    

Wavepackets are generated early in the simulations by adding a forcing term to the tendency of the difference $\delta$ in the divergence of the velocity field between  the two layers
\begin{eqnarray}
\frac{\partial}{\partial t} \delta = \ldots + A~ g_{10^\circ}(\theta - \pi/2)~ g_{\pi/m}(\phi - \omega t / m - \pi/m)~ \sin(\omega t - m \phi)
\label{wavepacket}
\end{eqnarray}
where $\ldots$ denotes the other terms in the equation of motion (Equation 5 of Ref. \cite{Marston:2012co}) and $g_{\Delta}(\alpha) \equiv \exp[-(\alpha / \Delta)^2]$ is a Gaussian of unit height centered at $\alpha = 0$ and of width $\Delta$.  The added driving acts for only half of an oscillation period, from $t = 0$ to $t = \pi / \omega$.  We choose $A = 10^{-4}$ (day)$^{-1}$, an amplitude sufficiently small for the waves to behave linearly.  Apart from the wave driving term at the early stage of simulation, all other forcing terms and friction are are turned off for the numerical experiments performed here.

An intermediate frequency wave train that consists largely of an eastward moving Kelvin wave is produced by setting the angular frequency $\omega = 1$ (day)$^{-1}$ and wavenumber $m = 10$ [Figure \ref{scattering}(a)].  A low frequency, long wavelength Kelvin wave is created by choosing instead $\omega = 0.2$ (day)$^{-1}$ and $m = 2$ [Figure \ref{scattering}(d)].  (Yanai waves can be generated by modifying the forcing term in Eq. \ref{wavepacket} to be antisymmetric with respect to the equator.)\footnote{Videos of the two simulations may be viewed at URLs https://vimeo.com/204597701 and https://vimeo.com/204598091.}

The localized region of enhanced stability breaks translational symmetry; therefore wavenumber is not conserved. Because the perturbation is static in time, however, the frequencies of the waves passing through the region remain unchanged. The intermediate frequency Kelvin wave packet propagates at a frequency that lies in the gap between the Rossby and Poincar\'e waves as indicated in Figure 1a in the Main Text.  There are thus no available scattering channels that match its frequency, and it emerges from the interaction intact [Figure \ref{scattering}(b, c)].  (Reflection symmetry of the Kelvin wave about the equator blocks scattering into the Yanai wave.)  Because the waves in the frequency gap have a topological origin, the absence of scattering is a manifestation of topological protection.  By contrast there is a band of Rossby waves at low frequencies, and the low-frequency Kelvin wave is strongly disturbed by its interaction with the enhanced stratification [Figure \ref{scattering}(e)], producing Rossby waves.   At late times, the low-frequency Kelvin wave has largely been converted to Rossby waves moving in the opposite (westward) direction [Figure \ref{scattering}(f)].   



\end{document}